\documentclass[slac_one]{revtex4}
\usepackage{graphicx}
\usepackage{fancyhdr}
\usepackage{amsmath} 
\usepackage{bm}
\usepackage{amsxtra}
\usepackage{amssymb}
\usepackage{amsthm}
\usepackage{latexsym}
\usepackage{lscape}

\usepackage{epstopdf}
\usepackage{braket}
\usepackage{units}
\usepackage{subfigure}

\begin{document}

\title{The {\sc Majorana Demonstrator}}

%
\newcommand{\alberta}{Centre for Particle Physics, University of Alberta, Edmonton, AB, Canada}
\newcommand{\blhill}{Department of Physics, Black Hills State University, Spearfish, SD, USA}
\newcommand{\ITEP}{Institute for Theoretical and Experimental Physics, Moscow, Russia}
\newcommand{\JINR}{Joint Institute for Nuclear Research, Dubna, Russia}
\newcommand{\lbnl}{Nuclear Science Division, Lawrence Berkeley National Laboratory, Berkeley, CA, USA}
\newcommand{\lanl}{Los Alamos National Laboratory, Los Alamos, NM, USA}
\newcommand{\queens}{Department of Physics, Queen's University, 
Kingston, ON, Canada}
\newcommand{\uw}{Center for Experimental Nuclear Physics and Astrophysics, 
and Department of Physics, University of Washington, Seattle, WA, USA}
\newcommand{\uchic}{Department of Physics, University of Chicago, Chicago, IL, USA}
\newcommand{\unc}{Department of Physics and Astronomy, University of North Carolina, Chapel Hill, NC, USA}
\newcommand{\duke}{Department of Physics, Duke University, Durham, NC, USA}
\newcommand{\ncsu}{Department of Physics, North Carolina State University, Raleigh, NC, USA}
\newcommand{\ornl}{Oak Ridge National Laboratory, Oak Ridge, TN, USA}
\newcommand{\ou}{Research Center for Nuclear Physics and Department of Physics, Osaka University, Ibaraki, Osaka, Japan}
\newcommand{\pnnl}{Pacific Northwest National Laboratory, Richland, WA, USA}
\newcommand{\sdsmt}{South Dakota School of Mines and Technology, Rapid City, SD, USA}
\newcommand{\usc}{Department of Physics and Astronomy, University of South Carolina, Columbia, SC, USA}
\newcommand{\usd}{Department of Earth Science and Physics, University of South Dakota, Vermillion, SD, USA}
\newcommand{\ut}{Department of Physics and Astronomy, University of Tennessee, Knoxville, TN, USA}
\newcommand{\tunl}{Triangle Universities Nuclear Laboratory, Durham, NC, USA}


\affiliation{\blhill}
\affiliation{\duke}
\affiliation{\ITEP}
\affiliation{\JINR}
\affiliation{\lanl}
\affiliation{\lbnl}
\affiliation{\ncsu}
\affiliation{\ornl}
\affiliation{\ou}
\affiliation{\pnnl}
\affiliation{\sdsmt}
\affiliation{\tunl}
\affiliation{\alberta}
\affiliation{\uchic}
\affiliation{\unc}
\affiliation{\usc}
\affiliation{\usd}
\affiliation{\ut}
\affiliation{\uw}

\author{E.~Aguayo}\affiliation{\pnnl} 
\author{F.T.~Avignone~III}\affiliation{\usc}\affiliation{\ornl}
\author{H.O.~Back}\affiliation{\ncsu}\affiliation{\tunl} 
\author{A.S.~Barabash}\affiliation{\ITEP}
\author{M.~Bergevin}\affiliation{\lbnl} 
\author{F.E.~Bertrand}\affiliation{\ornl}
\author{M.~Boswell}\affiliation{\lanl} 
\author{V.~Brudanin}\affiliation{\JINR}
\author{M.~Busch}\affiliation{\duke}\affiliation{\tunl}	
\author{Y-D.~Chan}\affiliation{\lbnl}
\author{C.D.~Christofferson}\affiliation{\sdsmt} 
\author{J.I.~Collar}\affiliation{\uchic}
\author{D.C.~Combs}\affiliation{\ncsu}\affiliation{\tunl}  
\author{R.J.~Cooper}\affiliation{\ornl}
\author{R.~J.~Creswick}\affiliation{\usc}  
\author{J.A.~Detwiler}\affiliation{\lbnl}
\author{P.J.~Doe}\affiliation{\uw}
\author{Yu.~Efremenko}\affiliation{\ut}
\author{V.~Egorov}\affiliation{\JINR}
\author{H.~Ejiri}\affiliation{\ou}
\author{S.R.~Elliott}\affiliation{\lanl}
\author{J.~Esterline}\affiliation{\duke}\affiliation{\tunl}
\author{J.E.~Fast}\affiliation{\pnnl}
\author{N.~Fields}\affiliation{\uchic} 
\author{P.~Finnerty}\affiliation{\unc}\affiliation{\tunl}
\author{F.M.~Fraenkle}\affiliation{\unc}\affiliation{\tunl} 
\author{V.M.~Gehman}\affiliation{\lanl}
\author{G.K.~Giovanetti}\affiliation{\unc}\affiliation{\tunl}  
\author{M.P.~Green}\affiliation{\unc}\affiliation{\tunl}  
\author{V.E.~Guiseppe}\affiliation{\usd}	
\author{K.~Gusey}\affiliation{\JINR}
\author{A.L.~Hallin}\affiliation{\alberta}
\author{R.~Hazama}{\affiliation{\ou}
\author{R.~Henning}\affiliation{\unc}\affiliation{\tunl}
\author{A.~Hime}\affiliation{\lanl}
\author{E.W.~Hoppe}\affiliation{\pnnl}
\author{M.~Horton}\affiliation{\sdsmt} 
\author{S. Howard}\affiliation{\sdsmt}  
\author{M.A.~Howe}\affiliation{\unc}\affiliation{\tunl}
\author{R.A.~Johnson}\affiliation{\uw} 
\author{K.J.~Keeter}\affiliation{\blhill}
\author{M.E.~Keillor}\affiliation{\pnnl}
\author{C.~Keller}\affiliation{\usd}
\author{J.D.~Kephart}\affiliation{\pnnl} 
\author{M.F.~Kidd}\affiliation{\lanl}	
\author{A. Knecht}\affiliation{\uw}	
\author{O.~Kochetov}\affiliation{\JINR}
\author{S.I.~Konovalov}\affiliation{\ITEP}
\author{R.T.~Kouzes}\affiliation{\pnnl}
\author{B.D.~LaFerriere}\affiliation{\pnnl}   
\author{B.H.~LaRoque}\affiliation{\lanl}	
\author{J. Leon}\affiliation{\uw}	
\author{L.E.~Leviner}\affiliation{\ncsu}\affiliation{\tunl}
\author{J.C.~Loach}\affiliation{\lbnl}	
\author{S.~MacMullin}\affiliation{\unc}\affiliation{\tunl}
\author{M.G.~Marino}\affiliation{\uw}
\author{R.D.~Martin}\affiliation{\lbnl}	
\author{D.-M.~Mei}\affiliation{\usd}
\author{J. Merriman}\affiliation{\pnnl}   
\author{M.L.~Miller}\affiliation{\uw} 
\author{L.~Mizouni}\affiliation{\usc}\affiliation{\pnnl}  
\author{M.~Nomachi}\affiliation{\ou}
\author{J.L.~Orrell}\affiliation{\pnnl}
\author{N.R.~Overman}\affiliation{\pnnl}  
\author{D.G.~Phillips II}\affiliation{\unc}\affiliation{\tunl}  
\author{A.W.P.~Poon}\affiliation{\lbnl}
\author{G. Perumpilly}\affiliation{\usd}   
\author{G.~Prior}\affiliation{\lbnl} 
\author{D.C.~Radford}\affiliation{\ornl}
\author{K.~Rielage}\affiliation{\lanl}
\author{R.G.H.~Robertson}\affiliation{\uw}
\author{M.C.~Ronquest}\affiliation{\lanl}	
\author{A.G.~Schubert}\affiliation{\uw}
\author{T.~Shima}\affiliation{\ou}
\author{M.~Shirchenko}\affiliation{\JINR}
\author{K.J.~Snavely}\affiliation{\unc}\affiliation{\tunl}	
\author{V. Sobolev}\affiliation{\sdsmt}  
\author{D.~Steele}\affiliation{\lanl}	
\author{J.~Strain}\affiliation{\unc}\affiliation{\tunl}
\author{K.~Thomas}\affiliation{\usd}		
\author{V.~Timkin}\affiliation{\JINR}
\author{W.~Tornow}\affiliation{\duke}\affiliation{\tunl}
\author{I.~Vanyushin}\affiliation{\ITEP}
\author{R.L.~Varner}\affiliation{\ornl}  
\author{K.~Vetter}\altaffiliation{Alternate address: Department of Nuclear Engineering, University of California, Berkeley, CA, USA}\affiliation{\lbnl}
\author{K.~Vorren}\affiliation{\unc}\affiliation{\tunl} 
\author{J.F.~Wilkerson}\affiliation{\unc}\affiliation{\tunl}\affiliation{\ornl}    
\author{B.A. Wolfe}\affiliation{\uw}	
\author{E.~Yakushev}\affiliation{\JINR}
\author{A.R.~Young}\affiliation{\ncsu}\affiliation{\tunl}
\author{C.-H.~Yu}\affiliation{\ornl}
\author{V.~Yumatov}\affiliation{\ITEP}
\author{C.~Zhang}\affiliation{\usd}					
			
\collaboration{{\sc{Majorana}} Collaboration}
\noaffiliation

\begin{abstract}
A brief review of the history and neutrino physics of double beta decay is given. A description of the {\sc Majorana Demonstrator} research and development  program including background reduction techniques is presented in some detail. The application of point contact (PC) detectors to the experiment is discussed, including the effectiveness of pulse shape analysis. The predicted sensitivity of a PC detector array enriched to 86\% in $^{76}$Ge is given.
\end{abstract}

\maketitle

\thispagestyle{fancy}

\section{Introduction}
There are three very important open questions in neutrino physics that can best be addressed by next generation zero-neutrino double-beta ($0\nu\beta\beta$) decay experiments. Are neutrinos Majorana particles that differ from antineutrinos only by helicity? What is their mass-scale? Is lepton number conservation violated? While searches for double beta decay have been carried out steadily throughout many decades, it is a very interesting time to launch next generation experiments. Measurements of atmospheric, solar and reactor neutrino oscillation have revealed scenarios in which the effective Majorana mass of the electron neutrino could be larger than 0.05~eV. Recent developments in Ge-detector technology make the search for $0\nu\beta\beta$ decay of $^{76}$Ge at this scale now feasible. For recent comprehensive experimental and theoretical reviews see~\cite{Elliott2000, Elliott2004, Avignone2008}.

The most sensitive limits on $0\nu\beta\beta$-decay half lives have come from Ge detectors enriched in $^{76}$Ge, namely the Heidelberg-Moscow ($T^{0\nu}_{\nicefrac{1}{2}}(^{76}\textrm{Ge}) \ge 1.9 \times 10^{25}$~y)~\cite{Baudis1999} and IGEX ($T^{0\nu}_{\nicefrac{1}{2}}(^{76}\textrm{Ge}) \ge 1.6 \times 10^{25}$~y)~\cite{Aalseth2004}, experiments. These bounds imply that the effective Majorana mass of the electron neutrino, $m_{\beta\beta}$, defined below, is less than about 0.6~eV using recently published nuclear matrix elements~\cite{Simkovic2009}. However, a subset of the Heidelberg-Moscow Collaboration reanalyzed the data and claimed evidence of a peak at the total decay energy, 2039 keV, implying $0\nu\beta\beta$ decay ~\cite{KK2001, KK2004, KK2006}. While this has been neither confirmed nor widely accepted by the neutrino community~\cite{Aalseth2002, Zdesenko2002, Feruglio2002}, there is no clear proof that the observed peak is not an indication of $0\nu\beta\beta$ decay. The GERDA experiment~\cite{Abt} is a $^{76}$Ge experiment being constructed in the Laboratori Nazionali del Gran Sasso (LNGS), and will probably be first to test this claim. The {\sc Majorana Demonstrator}, described below, will also directly test this claim. CUORE-0, the first tower of CUORE~\cite{Ardito}, also under construction at LNGS, will search for the $0\nu\beta\beta$ decay of $^{130}$Te, and will have the capability of confirming the claim; however, a null result would not be able to refute the claim because of the uncertainty in the theoretical nuclear matrix elements. 

The GERDA and {\sc Majorana}~\cite{Aalseth2005} collaborations intend to eventually come together to construct a tonne-scale experiment. This proposed $^{76}$Ge experiment, as well as the CUORE $^{130}$Te experiment~\cite{Ardito}, the EXO $^{136}$Xe experiment~\cite{Danilov2005}, the SNO+ $^{150}$Nd experiment~\cite{Kraus}, the KamLAND-Zen $^{136}$Xe experiment~\cite{Koga}, and the SuperNEMO $^{82}$Se/$^{150}$Nd experiment~\cite{Arnold} are all eventually designed to reach a mass sensitivity of $m_{\beta\beta} \approx 0.05$ eV to probe the inverted neutrino-mass hierarchy. Here we describe the {\sc Majorana Demonstrator} Project, which is designed to determine the feasibility of reaching the background level of 1-count per tonne-year in the 4-keV region of interest at 2039 keV in a tonne-scale $^{76}$Ge experiment.

There are other constraints on the neutrino-mass scale, irrespective of their Majorana or Dirac nature. The Troitsk~\cite{Belesev1999} and Mainz~\cite{Kraus2005} $^3$H single $\beta$-decay experiments have placed upper limits of 2.2-eV on the mass of the electron neutrino. The KATRIN experiment, a greatly enlarged tritium $\beta$-decay experiment in preparation, is projected to have a sensitivity of 0.2 eV~\cite{Osipowicz2003}.

Recent astrophysical data are also very relevant in a discussion of neutrino mass. For example Barger {\it et al.}~\cite{Barger2004} set a limit on the sum of neutrino mass eigenvalues, $\sum \equiv m_1 + m_2 + m_3 \le 0.75$~eV (95\% C.L.) from the Sloan Digital Sky Survey~\cite{Tegmark2004}, the two degree Field Galaxy Red Shift Survey (2dFGRS)~\cite{Percival2001}, and the Wilkinson Microwave Anisotropy Probe (WMAP)~\cite{Bennett2003}, as well as other CMB experiments and data from the Hubble Space Telescope. Hannestad~\cite{Hannestad} used the WMAP and 2dFGRS data to derive the bound $\sum < 1.0$~eV (95\% C.L.) and concluded that these data alone could not rule out the evidence claimed in~\cite{KK2001, KK2004, KK2006}. For recent papers on the subject see~\cite{Barger2003} and the references therein. The constraint $\sum \le 0.75$~eV would imply that the lightest neutrino eigenstate mass is $\le$0.24~eV. On the other hand, a positive value of $\sum$, implying $m_1 \approx 0.17$~eV, might also be interpreted to mean that next generation $0\nu\beta\beta$-decay experiments could constitute a stringent test of lepton-number conservation, irrespective of the neutrino mass hierarchy or the level of CP-violation in the weak sector.

\section{Neutrino Physics and Neutrinoless Double-Beta Decay}
Neutrino oscillation data very strongly imply that there are three eigenstates that mix and have mass. The flavor eigenstates, $\ket{\nu_{e,\mu,\tau}}$, are connected to the mass eigenstates, $\ket{\nu_{1,2,3}}$, via a linear transformation parameterized by a mixing matrix $u_{lj}^L$:
\begin{equation}
\label{eq:nuSup}
\ket{\nu_l} = \sum_{j=1}^3 \left| u_{lj}^L \right| e^{i\delta_j} \ket{\nu_j}.
\end{equation}
In Eq.~\ref{eq:nuSup}, $l = e,\mu,\tau$, and the factor $e^{i\delta_j}$ is a CP phase, $\pm1$ for CP conservation. The decay rate for the $0\nu\beta\beta$-decay mode driven by the exchange of a massive Majorana neutrino is expressed in the following approximation:
\begin{equation}
(T_{\nicefrac{1}{2}}^{0\nu})^{-1} = G^{0\nu}(E_{0}, Z) \left| \frac{m_{\beta\beta}}{m_{e}} \right|^{2} \left| M_{f}^{0\nu} - (g_A/g_V)^2 M_{GT}^{0\nu} \right|^2,
\end{equation}
where $G^{0\nu}$ is a phase space factor including the couplings, $m_{\beta\beta}$ is the effective Majorana mass of the electron neutrino given below, $M_{f}^{0\nu}$ and $M_{GT}^{0\nu}$ are the Fermi and Gamow-Teller nuclear matrix elements respectively, and $g_{A}$ and $g_{V}$ are the relative axial vector and vector weak coupling constants, respectively. After multiplication by a diagonal matrix of Majorana phases, $m_{\beta\beta}$ is expressed in terms of the first row of the 3$\times$3 matrix of Eq.~\ref{eq:nuSup} as follows:
\begin{equation}
\label{eq:mbb}
m_{\beta\beta} = \Big\vert  \left(u_{e1}^{L}\right)^{2} m_{1} + \left(u_{e2}^{L}\right)^{2} m_{2} e^{i\phi_{2}} +  \left(u_{e3}^{L}\right)^{2} m_{3} e^{i(\phi_{3} + \delta)} \Big\vert .
\end{equation}
In Eq.~\ref{eq:mbb}, $e^{i\phi_{2,3}}$ are the Majorana CP phases ($\pm$1 for CP conservation in the lepton sector). These phases do not appear in neutrino oscillation expressions and hence are unobservable in oscillation experiments. Only the phase angle $\delta$ affects neutrino oscillation. Oscillation experiments have, however, constrained the mixing angles and thereby the $u_{lj}^{L}$ coefficients in Eq.~\ref{eq:mbb}. Expressing the coefficients in terms of the sins and cosines of the mixing angles and the mass-squared differences measured in solar ($\Delta m_S^2$) and atmospheric ($\Delta m_A^2$) neutrino oscillation experiments, Eq.~\ref{eq:mbb} in the inverted hierarchy case becomes:
\begin{equation}
m_{\beta\beta}^{in\nu} = \sqrt{m_{3}^{2} + \Delta m_{A}^{2}}\left(c_{12}^{2} c_{13}^{2}\right) \pm \sqrt{m_{3}^{2} + \Delta m_{S}^{2} + \Delta m_{A}^{2}}\left(s_{12}^{2} c_{13}^{2}\right) + m_{3} s_{13}^{2} , 
\end{equation}
where $c_{ij}^{2} \equiv \cos^{2}(\theta_{ij})$ and $s_{ij}^{2} \equiv \sin^{2}(\theta_{ij})$. It is straightforward to arrive at an approximate expression for the target sensitivity of an experiment that should be able to discover $0\nu\beta\beta$ decay in the case of the inverted neutrino hierarchy. First we choose the maximum value i.e., the (+) sign, and we set the lowest mass eigenstate, $m_{3} = 0$, and $\Delta m_{s}^{2} \ll \Delta m_{A}^{2}$. This immediately yields the target value for the Majorana mass of the electron neutrino as $m_{\beta\beta} \approx c_{13}^{2} \sqrt{\Delta m_{A}^{2}} \approx$ 0.050 eV. The design of an experiment that does not meet these goals will not probe the inverted hierarchy region. While the {\sc Majorana Demonstrator} is not designed to meet these goals, the tonne-scale experiment for which it is the R\&D is. In this analysis, for simplicity we have assumed that CP is conserved in the lepton sector.

Until the month of this meeting, it was commonly thought that $\theta_{13}$ could be zero or extremely small. Recently, however, both T2K~\cite{Abe2011} and MINOS~\cite{Adamson2011} reported data that strongly imply a non-zero $\theta_{13}$ at a $\sim$90\% confidence level. In the case of the inverted hierarchy, T2K reports  $0.04 \le \sin^{2}(2 \theta_{13}) \le 0.34$ when $\delta_{CP} = 0$, at 90\% level of confidence. On the other hand, MINOS reports $2 \sin^{2}(\theta_{23}) \sin^{2}(2 \theta_{13}) = 0.079_{-0.053}^{+0.079}$. The MINOS data imply that the $\theta_{13} = 0$ hypothesis is disfavored at 89\% C.L. A recent global fit to all oscillation data by Fogli {\it et al.}~\cite{Fogli2011} incorporating further indications particularly from solar and reactor neutrino experiments yields \textgreater 3$\sigma$ evidence for non-zero $\theta_{13}$.  Accordingly, in the case that $m_3 \neq 0$, the analysis above for the inverted hierarchy becomes somewhat more complicated. However, the qualitative results remain essentially the same.

\section{The Majorana Demonstrator}
The {\sc Majorana Demonstrator} Project is classified by the Department of Energy as a research and development project to establish the feasibility of building and operating an ultra-low background tonne-scale $^{76}$Ge $0\nu\beta\beta$-decay experiment with a background in the 4-keV region of interest of 1-count per tonne-year. The first phase of the project will involve building two to three cryostats containing about 40 kg of Ge detectors, and with up to 30 kg in detectors fabricated from Ge enriched to $>$86\% in $^{76}$Ge. 

The {\sc Majorana} Collaboration uses two main techniques for reducing radioactive background. One is the careful selection, screening, and fabrication of the detector materials for radiopurity, including electroforming copper cryostat parts from CuSO$_{4}$ solution. The other technique is pulse-shape analysis (PSA), which is very effective with special-made point-contact (PC) germanium detectors. In this application, PSA takes advantage of the fact that $\beta\beta$-decay events deposit their energy very locally in the detector (single-site events), while gamma rays that deposit enough energy to contribute to the background in the 2-MeV energy-region of interest typically scatter more than once and deposit their energy in several locations (multiple-site events). As the charges liberated in these energy depositions drift through the detector, the cluster multiplicity imprints itself in the detector pulses, which can then be analyzed to identify and remove those that deposited energy more than once at different locations in the detector. This technique was used in both the Heidelberg-Moscow and IGEX experiments; however, recently this technique has been greatly improved. 

The point-contact (PC) germanium detector was originally developed at Lawrence Berkeley National Laboratory~\cite{Luke1988} and further developed for dark matter and $\beta\beta$-decay searches at the University of Chicago~\cite{Barbeau2007}. They were later selected as the detector type for the {\sc Majorana Demonstrator}. A photograph of a typical PC detector crystal is shown in Fig.~\ref{fig:ppc}. The surface shown is the passivated ultra-high resistance surface with the point contact in the center. These detectors have a weighting potential that is strongly peaked, so that charges drifting in the detector give very little signal until they are in the immediate vicinity of the point contact. The contact geometry also results in very low fields throughout the bulk of the crystal, and hence very slow drift velocities and long drift times, as shown in the right hand side of Fig.~\ref{fig:ppc}. As a result, charges generated in different regions of the detector will give sharp pulses well-differentiated in time, facilitating their identification and discrimination. 

\begin{figure*}[ht]
\centering
\mbox{\subfigure{\includegraphics[width=60mm]{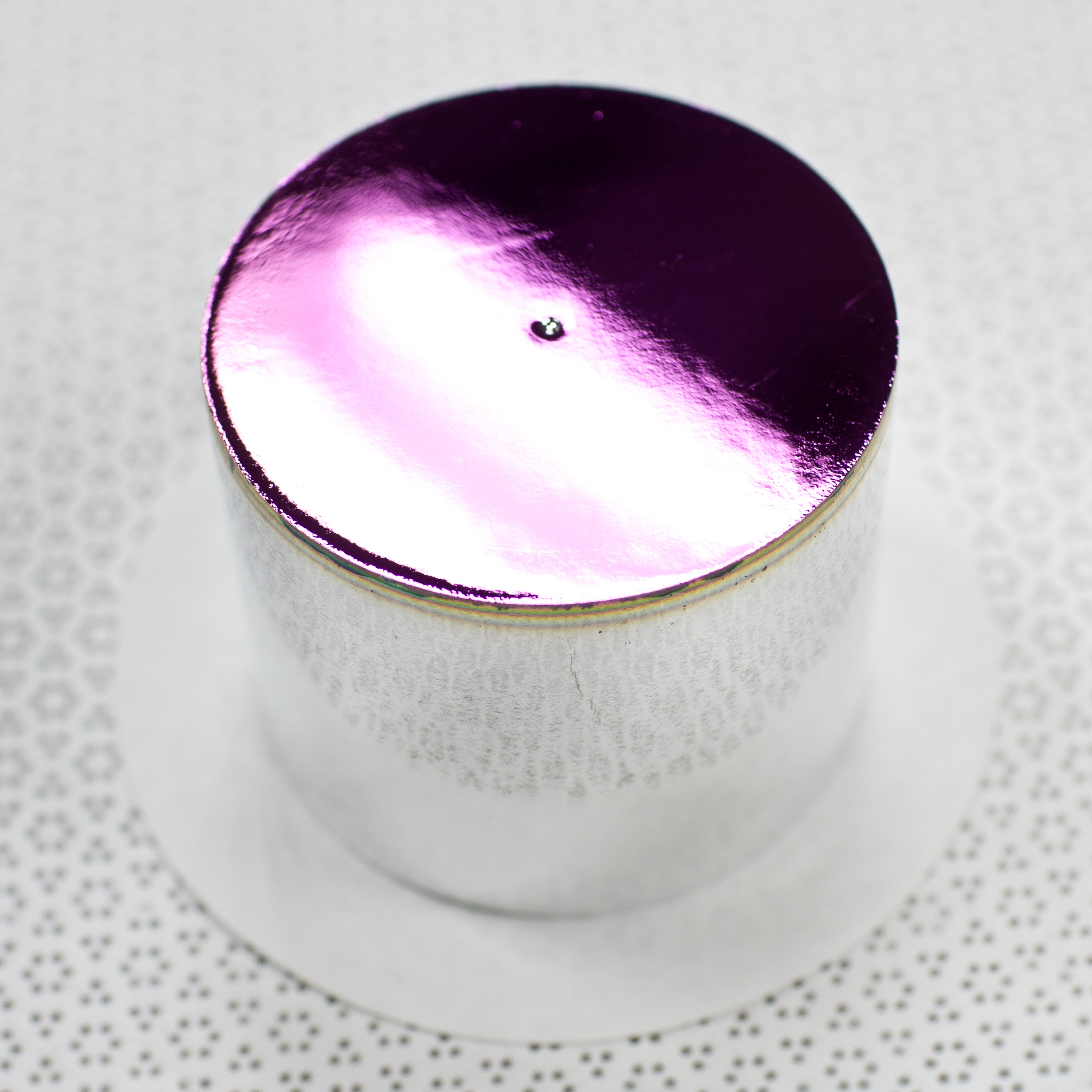}}\quad
\subfigure{\includegraphics[width=77mm]{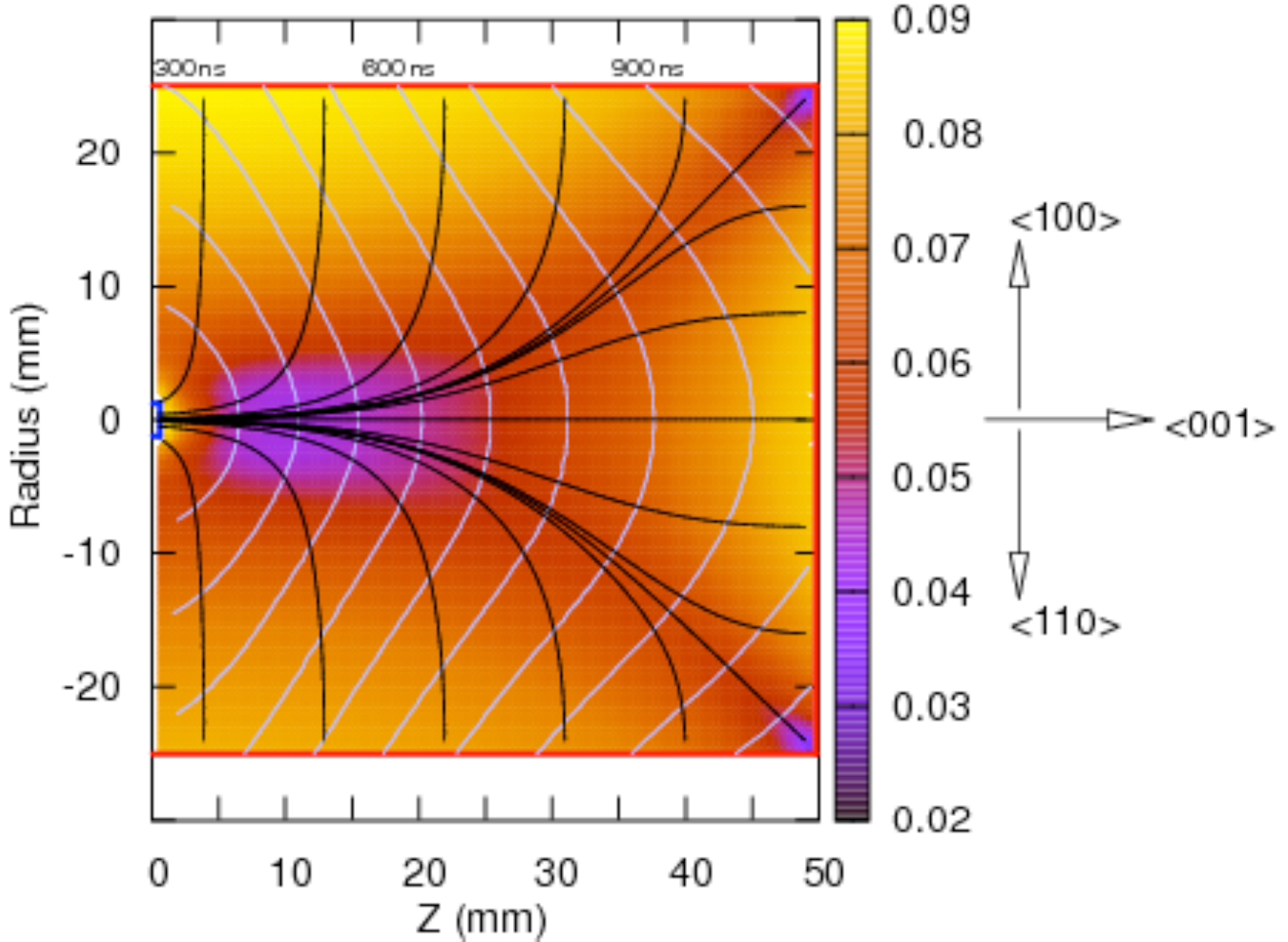} }}
\caption{(Left) A photograph of a point-contact detector showing the contact on the passivated surface. (Right) A diagram of a point-contact detector's electric field intensity (color pattern), drift trajectories (black lines), and isochrones (surfaces of constant drift time, grey).}
\label{fig:ppc}
\end{figure*}

A comparison of charge and current pulses from multi-site events in standard coaxial and PC HPGe detectors is shown in Fig.~\ref{fig:ppcPulses}. The result of applying PSA techniques to an actual experimental background spectrum is shown in Fig.~\ref{fig:psa}. The black curve is the best fit to the gamma ray lines in the $^{232}$Th decay chain from a calibration source. The lines at 1581, 1588, 1621, 1625, 1631, and 1638~keV are full-energy peaks corresponding to gamma rays of those energies, and therefore are expected to be dominated by multi-site events.  The peak at 1592-keV is the double-escape peak from the pair production interaction of the 2615-keV gamma-ray in $^{208}$Tl, the final daughter in the chain. The double-escape peak has a similar two-$\beta$ event topology and serves as a proxy for the $0\nu\beta\beta$-decay signal. The red curve is the best fit to the same spectrum after events with multi-site character have been removed. The full-energy gamma peak intensity is decreased by roughly an order-of-magnitude while the double-escape events remain with high efficiency. What is striking is the fact that even a majority of the events in the continuum from gamma-rays that partially deposited their energy, which are also dominantly multi-site in character, are recognized and removed. The conclusion from this experiment is that pulse shape analysis is a very effective background reduction technique with point-contact Ge detectors.

\begin{figure*}[ht]
\centering
\includegraphics[width=135mm]{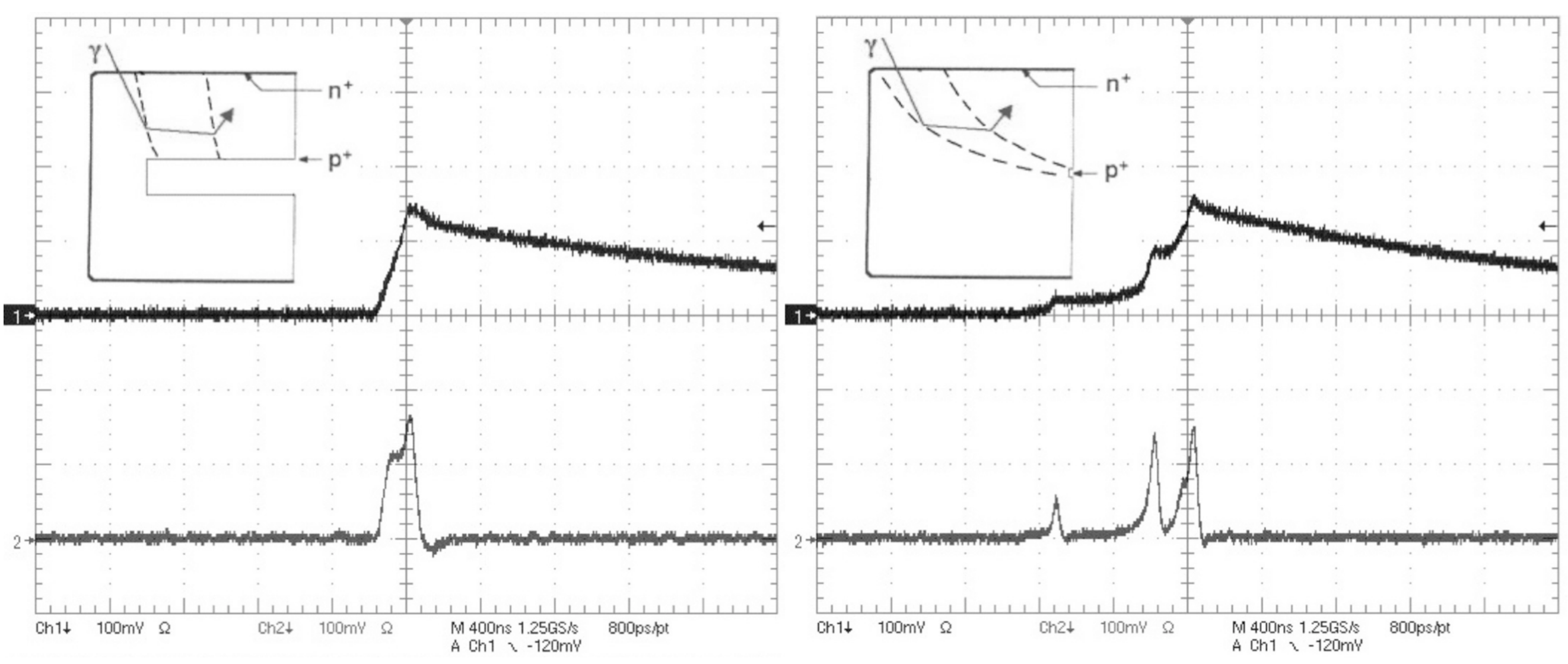}
\caption{(Left top) A typical charge pulse of an ordinary semi-coaxial high-purity germanium detector. (Left lower) The corresponding current pulse from differentiating the above pulse. (Right upper) A typical charge pulse from a point contact detector. (Right lower) the current pulse from differentiating the charging pulse above.}
\label{fig:ppcPulses}
\end{figure*}

\begin{figure*}[ht]
\centering
\includegraphics[width=135mm]{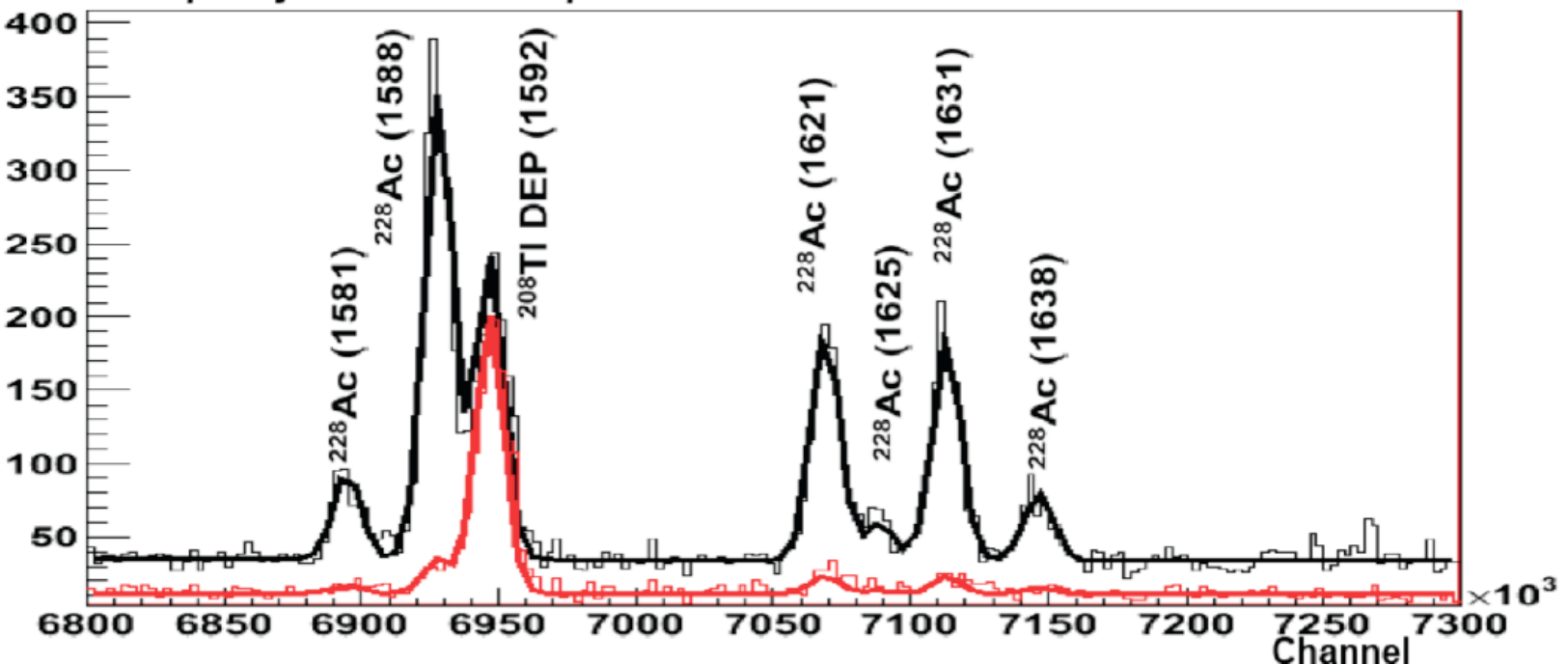}
\caption{A gamma-ray spectrum taken with a point-contact detector using a $^{232}$Th calibration source (Black fitted line). The lines at 1581, 1588, 1621, 1625, 1631, and 1638~keV are full-energy peaks corresponding to gamma rays of those energies, and are dominantly multi-site. The peak at 1592-keV is the double-escape peak from the 2615-keV line in the daughter $^{208}$Tl and serves as a proxy for the $0\nu\beta\beta$-decay signal. The red spectrum shows the events remaining after the application of PSA cuts to remove multi-site events.  A fit to the remnant peaks and background is also shown (red line).}
\label{fig:psa}
\end{figure*}

Figure~\ref{fig:MJD} shows a conceptual drawing of the {\sc Majorana Demonstrator} apparatus. Each cryostat is mounted in a self-contained module (monolith) in which the cross-arm containing the thermosyphon passes through a lead and electroformed copper shielding door. This assembly remains together if the module is removed and the cryostat opened. The monolith transporter is under construction at Los Alamos National Laboratory. With all of these ultra-low background techniques, the background goal for the {\sc Demonstrator} is 4-counts per tonne-year in the 4-keV region of interest. Monte-Carlo simulations were used to project that this would result in a background of $\sim$1-count per tonne-year in a one-ton close-packed array of detectors. Figure~\ref{fig:MJDSens} shows the predicted sensitivity~\cite{FeldmanCousins} versus exposure for of an array of germanium detectors enriched to 86\% in $^{76}$Ge using the nuclear matrix elements and phase space factors calculated in \cite{Simkovic2009} and \cite{G0nu}. The upper shaded portion is the range of the claimed observation reported in \cite{KK2006}. As can be seen in the figure, the {\sc Majorana Demonstrator} will be able to confirm or refute this claim with a $\sim$30~kg-y exposure. The lower shaded region corresponds to the range of $m_{\beta\beta}$ for the inverted hierarchy with $m_3 = 0$. The dark shading accounts for the variation of the unknown CP phases; the lighter shading includes the additional uncertainties in the neutrino oscillation parameters~\cite{PDG}. Our calculation reveals that with backgrounds below 1~count/ROI/t/y, a $0\nu\beta\beta$ decay search with $^{76}$Ge will be able to cover the entire inverted hierarchy with less than a 10 tonne-year exposure.

\begin{figure*}[ht]
\centering
\includegraphics[width=135mm]{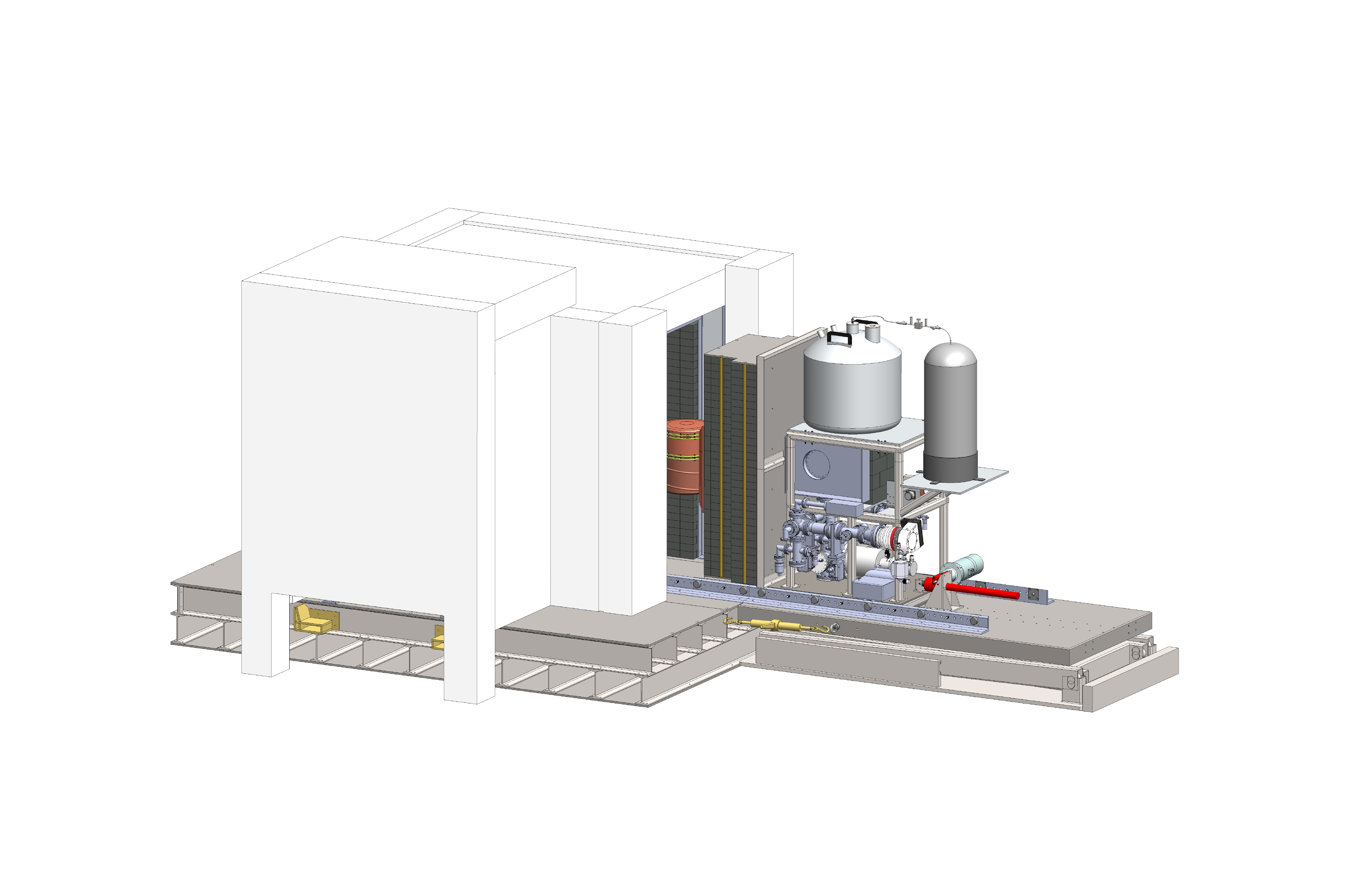}
\caption{An engineering concept drawing of two cryostat modules in the shield, with one partially withdrawn showing the copper cryostat.}
\label{fig:MJD}
\end{figure*}

\begin{figure*}[ht]
\centering
\includegraphics[width=135mm]{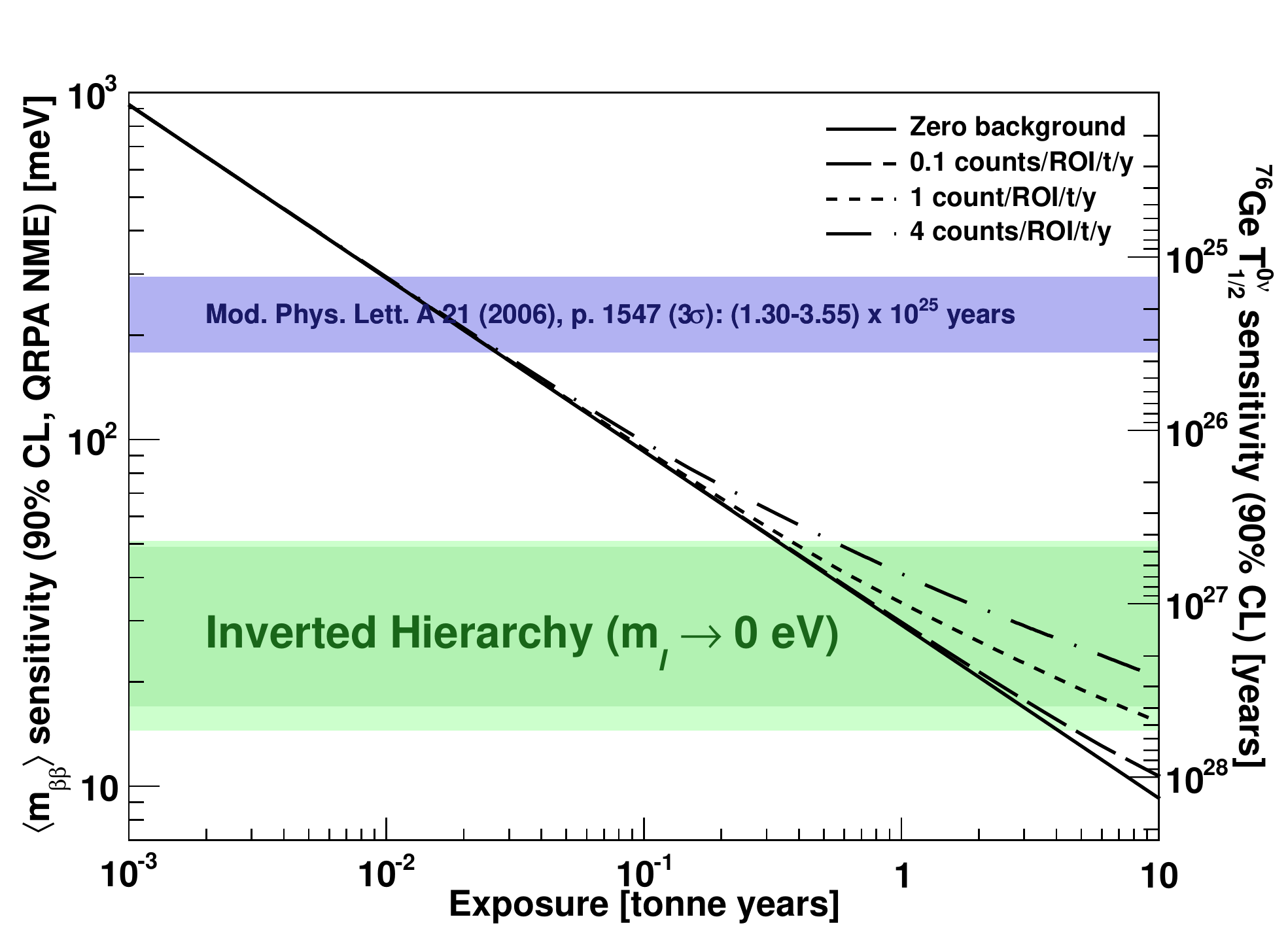}
\caption{The predicted sensitivity versus exposure of an array of germanium detectors enriched to 86\% in $^{76}$Ge}
\label{fig:MJDSens}
\end{figure*}

The {\sc Majorana} Collaboration was required to develop its own germanium purification facilities and processes for reducing the GeO$_2$ to metal and zone refining it to a resistance of 47~$\Omega$/cm. No experienced commercial company would reduce the enriched GeO$_2$ nor zone-refine it. A building was rented next to Electrochemical Systems Inc., (ESI) in Oak Ridge, Tennessee, and consultants formally in the Ge industry were hired. All the required equipment was purchased and installed. This portion of the {\sc Majorana Demonstrator} Project is supported by the National Science Foundation.

     The first step in the process is to heat the GeO$_2$ to about 650$^\circ$\,C in a hydrogen atmosphere in a graphite boat, which is in a quartz tube in a large furnace for about 9~hours. At this point the boat contains the Ge metal in a fine powder. Then the temperature is raised to 1030$^\circ$\,C for approximately 15~minutes to melt the powder to make a bar. When the Ge is completely melted the temperature is lowered, the atmosphere is changed to nitrogen, and the Ge cooled. After three R\&D runs, the remainder of the 29~kg of GeO$_2$ was reduced yielding 18.55~kg of metal with a resistivity of 47~$\Omega$/cm or better. The reduction of this quantity was accomplished with an efficiency of 98.9\%.

     This material was then zone refined in a graphite boat moving through a triple coil RF heater creating melted zones approximately a few cm in length. Each pass lasted about 11 hours. After two passes, the zone-refined bars were cooled and the resistivity measured. Portions of the bars that measured less than 47~$\Omega$/cm were cut off and zone refined again. The final product is 17~kg of detector grade Ge. A test detector will be fabricated as the final verification of the purification and material quality before the germanium enriched to $>$86\% in $^{76}$Ge will be subjected to these processes.

The first 29 kg of GeO$_{2}$, containing 20 kg of Ge metal enriched to $>$86\% in $^{76}$Ge in Russia, arrived in Oak Ridge, Tennessee on September 6 of this year. It will be stored in a shallow underground facility with an overburden of approximately 100 meters of water equivalent, to shield it from hard cosmic-ray generated neutrons that can produce spallation reactions in the Ge. These can generate backgrounds from $^3$H, $^{54}$Mn, $^{57}$Ni, $^{57,58,60}$Co, $^{65}$Zn, $^{67}$Ga and $^{68}$Ge. These isotopes begin to grow in immediately after the enrichment; however, all but $^{68}$Ge are eliminated by the Czochralski crystal-pulling process. At that time all of the isotopes begin to grow in at a rate depending on the overburden and altitude of their location. The logistics of our processing and detector production is designed to limit the exposure to cosmic rays to less than 100 days sea level equivalent. This exposure will satisfy our allowed background budget.

\begin{acknowledgments}
We acknowledge support from the Office of Nuclear Physics in the DOE
Office of Science under grant numbers DE-AC02-05CH11231, DE-FG02-97ER41041,
DE-FG02-97ER41033, DE-FG02-97ER4104, DE-FG02-97ER41042, DE-SCOO05054,
DE-FG02-10ER41715, and DE-FG02-97ER41020. We acknowledge support
from the Particle and Nuclear Astrophysics Program of the National Science
Foundation through grant numbers PHY-0919270, PHY-1003940, 0855314,
and 1003399. We gratefully acknowledge support from the Russian Federal
Agency for Atomic Energy. We gratefully acknowledge the support of the
U.S. Department of Energy through the LANL/LDRD Program. N. Fields
is supported by the DOE/NNSA SSGF program.
\end{acknowledgments}

\bigskip 


\begin{thebibliography}{99}

\bibitem{Elliott2000}
S.R Elliott and P. Vogel, Annu. Rev. Part. Sci. {\bf 52}, 115 (2000).
 
\bibitem{Elliott2004}
S.R. Elliott and J. Engel, J. Phys. G {\bf 30}, R183 (2004). 
 
\bibitem{Avignone2008} 
F.T. Avignone III, S.R. Elliott, and J. Engel, Rev. Mod. Phys. {\bf 80}, 481 (2008).
 
\bibitem{Baudis1999}
L. Baudis {\it et al.} (The Heidelberg-Moscow Collaboration), Phys. Rev. Lett. {\bf 83}, 41 (1999).

\bibitem{Aalseth2004}
C.E. Aalseth {\it et al.} (The IGEX Collaboration), Phys. Rev. C {\bf 59}, 2108 (1999); Phys. Rev. D {\bf 65}, 092007 (2002); Phys. Rev. D {\bf 70}, 078302 (2004).

\bibitem{Simkovic2009}
F. Simkovic {\it et al.}, Phys. Rev. C {\bf 79}, 055501 (2009).

\bibitem{KK2001}
H.V. Klapdor-Kleingrothaus, A. Deitz, H.L.Harney and I.V. Krivosheina, Mod. Phys. Lett. A {\bf 16}, 2409 (2001).

\bibitem{KK2004}
H.V. Klapdor-Kleingrothaus {\it et al.}, Phys. Lett. B {\bf 586}, 198 (2004); Nucl. Inst.  Meth. Phys. Res. A {\bf 522}, 371 (2004).

\bibitem{KK2006}
H.V. Klapdor-Kleingrothaus {\it et al.}, Mod. Phys. Lett. A {\bf 21}, 1547 (2006).

\bibitem{Aalseth2002}
C.E. Aalseth {\it et al.}, Mod. Phys. Lett. A {\bf 17}, 1475 (2002).

\bibitem{Zdesenko2002} 
Yu. G. Zdesenko, F.A. Danevich, and V.I. Tretyak, Phys. Lett. B {\bf 546}, 206 (2002). 

\bibitem{Feruglio2002}
F. Feruglio, A. Strumia and F. Vissani, Nucl. Phys. B {\bf 637}, 345 (2002). 

\bibitem{Abt}
I. Abt {\it et al.} (The GERDA Collaboration), arXiv:hep-ex/0404039 (2004).

\bibitem{Ardito}
R. Ardito {\it et al.} (The CUORE Collaboration), arXiv:hep-ex/0501010 (2005).

\bibitem{Aalseth2005}
C. E. Aalseth {\it et al.}, (The Majorana Collaboration), Nucl. Phys. B (Proc. Suppl.) {\bf 138}, 217 (2005); Also see arXiv:nucl-ex/0311013.

\bibitem{Danilov2005}
M. Danilov {\it et al.}, Phys. Lett. B {\bf 480}, 12 (2000); D. Akimov {\it et al.}, Nucl. Phys. B Proc. Suppl. {\bf 138}, 224 (2005).

\bibitem{Kraus}
C. Kraus and S.J. M. Peeters, Prog. Part. Nucl. Phys. {\bf 64}, 273 (2010).

\bibitem{Koga}
M. Koga, ``KamLAND double betadecay experiment using $^{136}$Xe,'' International Conference on High Energy Physics (ICHEP), Paris, France (July  2010).

\bibitem{Arnold}
R. Arnold {\it et al.} (The SuperNEMO Collaboration), Eur. Phys. J. C {\bf 70}, 927 (2010); R. Saakyan {\it et al.}, L. Phys. Conf. Ser. {\bf 179}, 101006 (2009).

\bibitem{Belesev1999}
A.I. Belesev {\it et al.}, Phys. Lett. B {\bf 350}, 263 (1995); V.M. Lobashev {\it et al.}, Phys. Lett. B {\bf 460}, 227 (1999).

\bibitem{Kraus2005}
Ch. Kraus {\it et al.}, Eur. Phys. J. C {\bf 40}, 447 (2005).

\bibitem{Osipowicz2003}
A. Osipowicz {\it et al.}, arXiv:hep-ex/0109033 (2001); V.M. Lobashev, Nucl.Phys. A {\bf 719}, 153 (2003), and refs. therein.

\bibitem{Barger2004}
V. Barger {\it et al.}, Phys. Lett. B {\bf 595}, 55 (2004).

\bibitem{Tegmark2004}
M. Tegmark {\it et al.}, Phys. Rev. D {\bf 69}, 103501 (2004).

\bibitem{Percival2001}
W.J. Percival {\it et al.}, Mon. Not. Roy. Astron. Soc. {\bf 327}, 1297 (2001); M. Colless {\it et al.}, Mon. Not. Roy. Astron. Soc. {\bf 328},1039 (2001). 

\bibitem{Bennett2003}
C.L. Bennett {\it et al.}, Astrophys. J. Suppl. Ser. {\bf 148}, 1 (2003); D.N. Spergel {\it et al.}, Astrophys. J. Suppl. Ser. {\bf 148},175 (2003).

\bibitem{Hannestad}
S. Hannestad, JCAP {\bf 0305}, 004 (2003). 

\bibitem{Barger2003}
V. Barger, D. Marfatia and K. Whisnant, Int. J. Mod. Phys. E {\bf 12}, 569 (2003); see also P. Crotty, J. Lesgourgues, and S. Pastor, Phys. Rev. D {\bf 69}, 123007 (2004) and references therein.

\bibitem{Abe2011}
K. Abe, {\it et al.}, (The K2K Collaboration) arXiv:1106.2822 [hep-ex] (2011).

\bibitem{Adamson2011}
P. Adamson {\it et al.}, (The MINOS Collaboration) arXiv: 1108.0015 [hep-ex] (2011).

\bibitem{Fogli2011} 
G. L. Fogli, E. Lisi, A. Marrone, A. Palazzo, and A. M. Rotunno, arXiv:1106.6028v1 [hep-ph] (2011).

\bibitem{Luke1988}
P. N. Luke, N.W. Madden and F.S. Goulding, IEEE Trans. Nucl. Sci. {\bf 32}, 457 (1985); P. N. Luke Nucl. Instr. and Meth. A {\bf 271}, 567 (1988).

\bibitem{Barbeau2007}
P. S. Barbeau, J. I. Collar and O. Tench, JCAP {\bf 0709}, 009 (2007). 

\bibitem{FeldmanCousins}
G. J. Feldman and R. D. Cousins, Phys. Rev. D {\bf 57}, 3873 (1998).

\bibitem{G0nu}
J. Suhonen and O. Civitarese, Phys. Rep. {\bf 300}, 124 (1998).

\bibitem{PDG}
K. Nakamura {\it et al.} (Particle Data Group), J. Phys. G {\bf 37}, 075021 (2010).

\end{thebibliography}
\end{document}